\documentclass{IET-Conf-Paper}
\usepackage{tikz}
\usetikzlibrary{arrows.meta,positioning,decorations.pathreplacing}
\usepackage{subcaption} 
\usepackage{booktabs}
\usepackage{multirow}
\usepackage{etoolbox}
\usepackage{blindtext}
\usepackage{soul}
\usepackage{lipsum}
\usepackage[hidelinks]{hyperref}  
\makeatletter
\patchcmd{\@afterheading}{\@afterindentfalse}{\@afterindenttrue}{}{}
\makeatother

\header{}

\begin{document}

\title{UNDERSTANDING RISK AND REVENUE IN THE NORDIC 15-MINUTE MFRR MARKET: AN EV AGGREGATION STUDY}

\author{Theodor Hagström\ad{1,2}\corr, Lars Herre\ad{1,2,3}}

\address{%
  \add{1}{Fortum AB, Solna, Sweden}
  \add{2}{Hiven Energy, Solna, Sweden}
  \add{3}{DTU Technical University of Denmark, Kgs. Lyngby, Denmark}
{\email{theodor.hagstrom@fortum.com}}
}

\keywords{MANUAL FREQUENCY RESTORATION RESERVE, 15-MINUTE MARKET TIME UNIT, ELECTRIC VEHICLE AGGREGATOR, BIDDING STRATEGY, STOCHASTIC OPTIMISATION}

\begin{abstract}
Decarbonisation, decentralisation, and intermittency are driving the development of flexibility markets towards shorter market time units (MTU). Shorter MTUs and shorter gate closures lower the entrance barriers of demand side aggregators that face significant uncertainty on longer time scales. We study the business case for aggregated EV fleets participating in the Nordic 15-minute mFRR Energy Activation Market~(EAM). Motivated by increasing system granularity and rapid EV uptake, we represent fleet flexibility as a \textit{virtual battery} with time-varying power and energy envelopes and formulate a risk-aware stochastic optimisation that co-ordinates day-ahead scheduling with quarter-hour mFRR bidding. Using synthetic residential charging cohorts and observed day-ahead prices on two stylised days, we compare an \textit{independent} day-ahead baseline to a \textit{co-optimised} strategy under conservative availability and a CVaR-augmented objective. Across both price cases, co-optimisation increases expected profit and lowers downside risk: the model buys less energy day-ahead and shifts procurement toward mFRR~down while flattening the charging plan to retain eligibility for mFRR~up. Profit decomposition shows that the uplift is driven by higher mFRR~down revenues and reduced reliance on unwinding day-ahead positions. We discuss operational implications for bidding and outline two extensions: rolling 45-minute re-optimisation and a V2G framework.
\end{abstract}

\maketitle

\section{Introduction}
\subsection{Context}

Growing shares of variable renewables increase short-term forecast errors and net-load volatility, which raises the need for balancing products and for market designs that align price signals with physical imbalances at finer granularity. In the Nordics, this has driven a shift to quarter-hour resolution: the mFRR Energy Activation Market (EAM) already clears in 15-minute Market Time Units (MTUs) as of 4~March~2025, and the day-ahead market is also moving to 15-minute MTUs from 30~September~2025~\cite{nordpoolfifteen}.

At the same time, electric-vehicle (EV) adoption continues to accelerate. In 2024, global electric car sales exceeded 17~million, surpassing a 20\% sales share; China accounted for almost half of all car sales and now has roughly one in ten cars on the road electric. Europe held around a 20\% sales share amid uneven national trends, while the United States reached just over one in ten new cars sold~\cite{iea2025evoutlook}. This rapid uptake is turning EVs into a major source of flexible demand with growing relevance for power system balancing.

In this study, we investigate how aggregated EV fleets can participate in the Nordic 15-minute mFRR EAM. Using an aggregation framework to represent fleet flexibility and a risk-aware stochastic optimisation model, we evaluate day-ahead scheduling strategies and short-term reserve bidding. The focus is on understanding the business case for aggregators and on analysing how expected revenues, risks, and profit drivers vary across market conditions.
\subsection{Related Work and Positioning}

The literature on reserve bidding spans several perspectives, from conventional generators to emerging distributed resources (see~\cite{ackermann2001} for definitions and terminology). Early contributions focused on both thermal and hydro units. Reference~\cite{anderson2002} studies optimal offer construction under demand and competitor uncertainty for thermal plants, Reference~\cite{boomsa2014} analyses coordinated bidding across sequential Nordic electricity markets, Reference~\cite{lohndorf2013} applies approximate dynamic programming to integrate scheduling and bidding for hydro storage, and Reference~\cite{khodadadi2021} examines coordinated hydropower bidding in day-ahead and mFRR markets. These studies focus on centralised, dispatchable units with tight operational control and well-characterised dynamics. This setting is quite different from pooling many small, decentralised EVs via an aggregator. Within this stream, Reference~\cite{boomsa2014} demonstrates the theoretical gain from explicitly co-optimising day-ahead and reserve bidding, compared to treating the two markets separately.

More recently, attention has shifted to aggregators of distributed loads, whose challenges differ from hydro-based models due to distributed ownership, uncertain availability, and charging constraints. In one of the earliest works, Reference~\cite{bessa2011} formulates three optimisation problems for aggregator participation in day-ahead and secondary reserve markets, capturing uncertainty in vehicle availability and charging demand. Reference~\cite{vagropoulos2013} extends this line by introducing a two-stage stochastic program with synthetic EV data to identify flexible charging periods, contrasting outcomes under perfect and naive forecasts. Risk aversion has been explicitly incorporated by Reference \cite{momber2015}, who applies a CVaR-augmented objective to a two-stage stochastic formulation, demonstrating how risk preferences influence bidding behaviour and charging cost recovery. Building on this, Reference~\cite{herre2019} studies the Nordic market and formulates a risk-averse two-stage model for day-ahead energy and frequency containment reserves, using real EV charging data. 

Despite this progress, important gaps remain. To the best of our knowledge, no study has evaluated the business case for EV aggregator participation in the new Nordic 15-minute mFRR Energy Activation Market (EAM), which was only introduced in March 2025. Prior work largely considers day-ahead or hourly scheduling horizons and focuses on capacity markets, while short-notice quarter-hour (QH) balancing has not been addressed in detail. 

Moreover, most contributions stop at formulating and solving the optimisation problem, without decomposing how profits are generated or which factors drive risk exposure. Finally, although Reference~\cite{boomsa2014} establishes the theoretical benefit of co-optimisation, there has been little qualitative analysis of how independent and co-optimised bidding strategies differ in practice for EV fleets. This paper addresses these gaps by examining the Nordic 15-minute mFRR EAM with a virtual-battery aggregation framework, analysing the business case in terms of profit sources and risks, and providing a comparative evaluation of independent versus co-optimised bidding.

Specifically, our contributions are the following:
\begin{itemize}
    \item We co-optimise the scheduling of an EV aggregator including the novel 15-minute mFRR Energy Activation Market (EAM).
    \item We decompose the results and investigate drivers for risk and expected revenue.
    \item We analyse how different bidding strategies differ in practice for EV fleets.
\end{itemize}

\subsection{Article Outline}
The rest of the paper proceeds as follows. Section \ref{sec:marketRules} explains the market setting and rules. Section \ref{sec:methodology} introduces the EV aggregation as a virtual battery with power and energy envelopes, defines the risk-aware optimisation problem and bidding logic, and explains the scenario design and case-study set-up. Section~\ref{sec:results} reports key performance indicators, illustrates charging patterns, activations and prices along a representative scenario path, and decomposes profits to clarify revenue sources and risks. Section~\ref{sec:discussion} interprets operational implications for bidding, provides a practical participation guide for the mFRR market, and outlines extensions towards V2G. Section~\ref{sec:conclusion} provides practical conclusions for EV aggregators aiming to participate in quarter-hourly energy-rich reserve markets.

\section{Market Context and Rules}\label{sec:marketRules}
This section summarises the Nordic mFRR design and operation: the capacity and energy activation markets, timelines and activation requirements, and pricing and settlement.

\subsection{The Nordic mFRR Market}
The Manual Frequency Restoration Reserve (mFRR) provides balancing energy and is procured by the Transmission System Operator (TSO) in two separate markets. In the capacity market (mFRR~CM), which closes at 07:30 on day-ahead (D–1), providers are compensated for the capacity they commit. By contrast, the energy activation market closes 45~minutes before delivery (to be shortened to 25~minutes with the accession to MARI~\citep{nbm2024mfrr}) and compensates providers only for activated bids. Both markets are structured into separate \textit{up} and \textit{down} products. For a consumption portfolio, mFRR~up implies reducing charging (lower consumption corresponds to \textit{up}-regulation of generation), whereas mFRR~down implies increasing charging (higher consumption corresponds to \textit{down}-regulation of generation).

\begin{figure}[h]
  \centering
  \resizebox{\columnwidth}{!}{

\begin{tikzpicture}[
  x=0.18cm, y=1.7cm, >=Latex,
  tick/.style={thin},
  labelbelow/.style={font=\footnotesize,align=center,anchor=north,yshift=-2pt,text width=22mm},
  labelabove/.style={font=\footnotesize,align=center,anchor=south,yshift=2pt,text width=22mm},
  note/.style={font=\scriptsize,align=center},
  brace/.style={decorate,decoration={brace,amplitude=4pt}}
]

\pgfmathsetmacro{\GC}{-30}      
\pgfmathsetmacro{\NOTIFY}{-7.5} 
\pgfmathsetmacro{\QS}{0}        
\pgfmathsetmacro{\QE}{15}       

\pgfmathsetmacro{\HOLD}{2.5}                 
\pgfmathsetmacro{\RAMP}{10}                  
\pgfmathsetmacro{\Hend}{\NOTIFY+\HOLD}       
\pgfmathsetmacro{\Rend}{\Hend+\RAMP}         
\pgfmathsetmacro{\FULL}{\Rend}               
\pgfmathsetmacro{\Fend}{\Rend + 5}           
\pgfmathsetmacro{\Aend}{\Fend + \RAMP}       

\draw[thin] (\GC,0) -- (\Aend,0);
\foreach \x in {\GC,\NOTIFY,\QS,\Rend,\Rend+5,\QE} {\draw[tick] (\x,0.12) -- (\x,-0.12);}

\tikzset{qhlabel/.style={labelbelow, yshift=-4pt, rotate=-15}}

\node[labelbelow] at (\GC,0.5)     {Gate closure};
\node[qhlabel] at (\GC,0)     {QH--45};
\node[labelbelow] at (\NOTIFY,0.5) {Notification};
\node[qhlabel] at (\NOTIFY,0) {QH--7.5};
\node[qhlabel] at (\QS,0)     {QH--0};
\node[labelabove] at (\Rend+2.5,1.1)   {Full activation};
\node[qhlabel] at (\Rend,0)   {QH+5};
\node[qhlabel] at (\Rend+5,0) {QH+10};
\node[qhlabel] at (\QE,0)     {QH+15};

\draw[thick] (\NOTIFY,0) -- (\Hend,0) -- (\Rend,1) -- (\Fend,1) -- (\Aend,0);

\fill[gray!20] (\Hend,0) -- (\Rend,1) -- (\Rend,0) -- cycle; 
\fill[gray!20] (\Rend,0) rectangle (\Fend,1);                  
\fill[gray!20] (\Fend,1) -- (\Aend,0) -- (\Fend,0) -- cycle;   

\pgfmathsetmacro{\MIDHOLD}{0.5*(\NOTIFY+\Hend)}
\pgfmathsetmacro{\MIDRAMP}{0.5*(\Hend+\Rend)}
\pgfmathsetmacro{\MIDSTEADY}{0.5*(\Rend+\QE)}
\node[note,anchor=south]      at (\MIDRAMP,1.02) {};
\node[note,anchor=south]      at (\MIDSTEADY,1.02) {};

\draw[red, dashed, thick] (\QS,0) -- (\QS,1);
\draw[red, dashed, thick] (\QE,0) -- (\QE,1);

\end{tikzpicture}}
  \caption{The quarter-hour timeline for mFRR: gate closure (current QH--45, transitioning to QH--25), notification, ramp up, full activation, and ramp down. Start and end of the operating quarter-hour in red, dashed lines. }
  \label{fig:mfrr-timeline}
\end{figure}
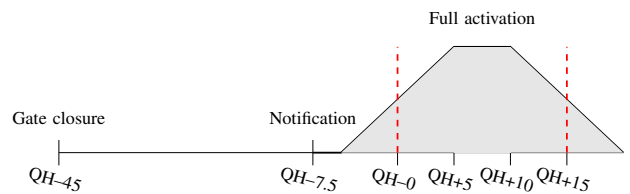

The TSO sends an activation signal 7.5~minutes before the operating period, and requires participants to reach full activation within 12.5~minutes. This makes mFRR a \textit{slow} ancillary service compared with FRR/FCR/aFRR. It also means that mFRR can be provided by direct steering of the EV charging, without requiring steering via the charger. Fig.~\ref{fig:mfrr-timeline} illustrates the \textit{scheduled activation} mFRR EAM product. Ramping is required to be symmetrical around the QH--0 and QH+15 mark. 

Activated bids are settled at the marginal activation price for the quarter-hour, under a uniform pricing scheme. By definition, mFRR~up prices are greater than or equal to, and mFRR~down prices less than or equal to, the day-ahead spot price for the same time period. We refer to the deviation from the day-ahead price as the \textit{mFRR premium}. Balancing Service Providers (BSPs) may submit \textit{simple} bids, as single price–volume pairs, or combine them into \textit{complex} piecewise offer curves. Bid volumes must be integer values in MW, with a minimum of 1~MW.

In March~2025, the Nordic region launched the mFRR Energy Activation Market (EAM), introducing 15-minute MTUs for bidding and pricing. The gate closure time is 45 minutes before the respective quarter-hour, and an mFRR bid is accepted, that is, activated, 7.5~minutes before the start of the quarter-hour. Therefore, placed bids are still pending for acceptance or rejection at the time that the next bids must be submitted for the consecutive quarter-hour. To accommodate this, the market allows limited bid linking across time periods. A technical link ties a bid to one activated in the immediately preceding quarter-hour, while a conditional link (available only for simple bids) can extend up to two quarter-hours backward and cover at most six bids in total~\citep{entsoe2023mari}.

\subsection{Imbalance Settlement and BRP Exposure}
The Nordic area applies a one-price imbalance settlement model at 15~minute resolution: deviations are settled at the quarter-hour’s marginal mFRR activation price, regardless of whether the deviation helps or worsens the system imbalance.  This implies that a Balance Responsible Party (BRP) can profit from being in imbalance when it supports the power system, while it is penalised for imbalances that worsen the power system balance in the respective MTU. Small fixed fees are also added regardless of direction. If the aggregator is also the BRP, this exposure is direct; otherwise it is typically passed through contractually, with the aggregator bearing financial responsibility.

\section{Methodology}\label{sec:methodology}
This section sets out the EV aggregation framework and data assumptions, introduces the virtual-battery representation, defines the risk-aware optimisation and bidding set-up, and explains the scenario design and case study.

\subsection{Aggregator Data Readiness and Vehicle Aggregation}\label{sec:EV}

\subsubsection{EV Uncertainty}\label{sec:ev-uncertainty}
Aggregating and bidding with distributed EVs poses two main challenges. The aggregator must meet the individual vehicle requirements while also fulfilling commitments to the TSO. At the time of bid submission, the exact state of the EV fleet at the time of delivery is unknown. This uncertainty is material in markets with day-ahead gate closure, like FCR, aFRR, and mFRR capacity markets. For the mFRR energy activation market, which closes 45~minutes before delivery, the uncertainty is lower but not negligible.

At 45~minutes before delivery, the aggregator typically knows, for each plugged-in vehicle, its remaining energy demand (kWh), charging peak power (kW), and desired departure time. Remaining uncertainty concerns future arrivals and potential early disconnects (e.g., unannounced departures or connectivity issues). To fully utilise flexibility, these events can be forecast from historical patterns and incorporated in decisions. As forecasting is out of scope in this work, a conservative approach is employed which bases bids only on vehicles that are currently plugged in and not scheduled to leave. Furthermore, we include a safety margin (e.g., 10~\%) for unexpected unavailability. For example, a 1~MW bid would require roughly 100~EVs at 11~kW, each connected and not due to depart within the next hour to cover the bid size, while reserving a buffer capacity.

\subsubsection{Virtual-Battery Representation}
Managing a large, distributed fleet vehicle-by-vehicle is impractical even with the above precautions. Computation should scale approximately independently of fleet size, whereas enforcing each vehicle’s constraints one-by-one quickly leads to prohibitive computation times.

A practical alternative is a \textit{virtual battery} that aggregates individual charging needs into a single controllable resource~\citep{brinkel2023}. Each vehicle can be described by four attributes: energy required to reach its target state of charge, arrival time, departure time, and maximum charging power. Aggregation preserves these requirements without representing every car individually in the optimiser.

The virtual battery is defined by three time-varying envelopes:
\begin{itemize}
  \item $E_t^{\min}$: \textit{latest-start} bound -- minimum energy required by time $t$ to ensure all departures leave fully charged.
  \item $E_t^{\max}$: \textit{earliest-start} bound -- maximum energy that could have been charged by time $t$ if all EVs charged immediately upon arrival.
  \item $P_t^{\max}$: instantaneous charging capacity -- the sum of all chargers' peak power that are plugged in at time $t$.
\end{itemize}

\begin{figure}[h]
  \centering
  \includegraphics[width=\columnwidth]{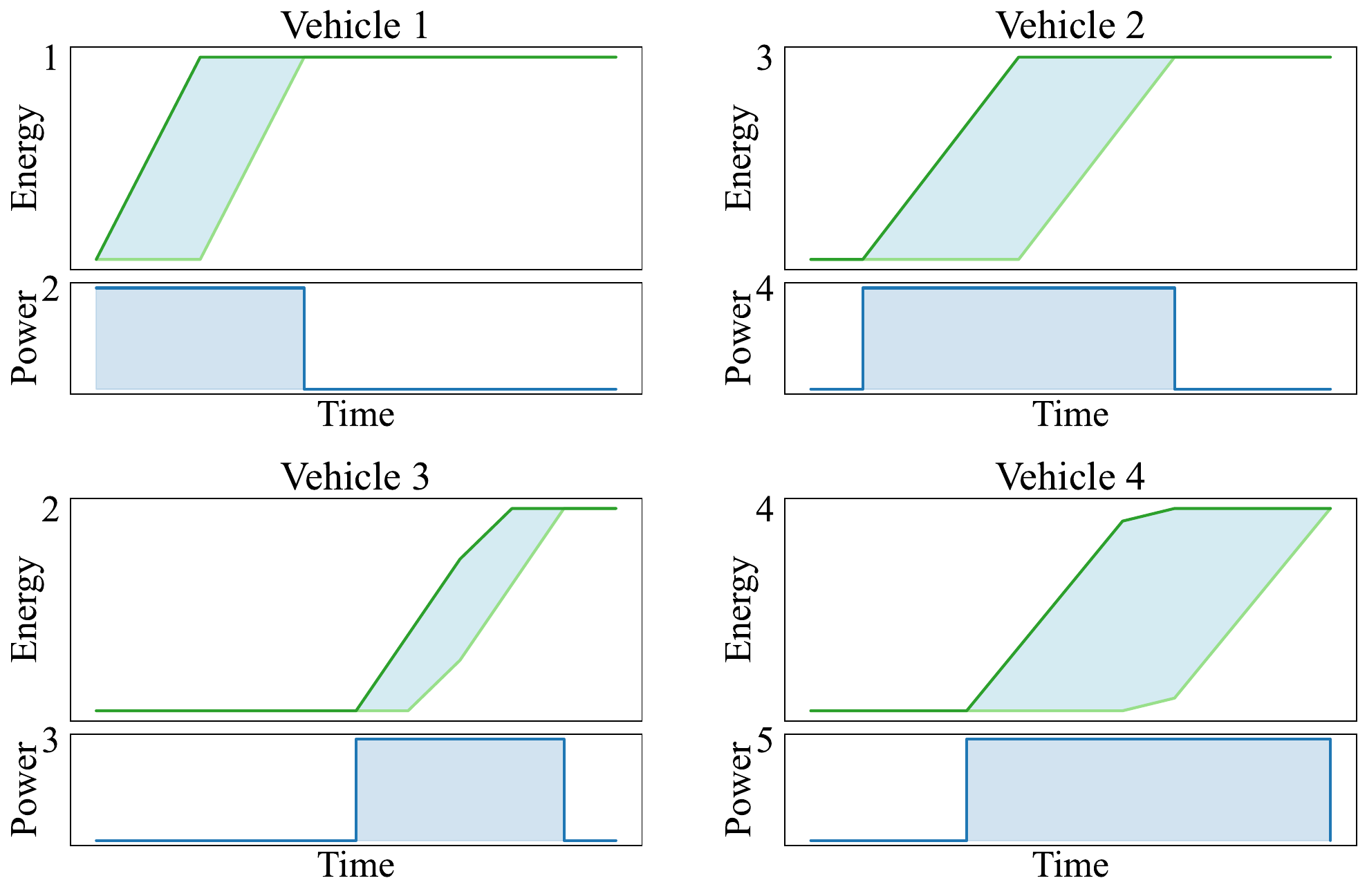}
  {\\ \footnotesize (a) Individual schedules\par}

  \vspace{4pt}

  \includegraphics[width=\columnwidth]{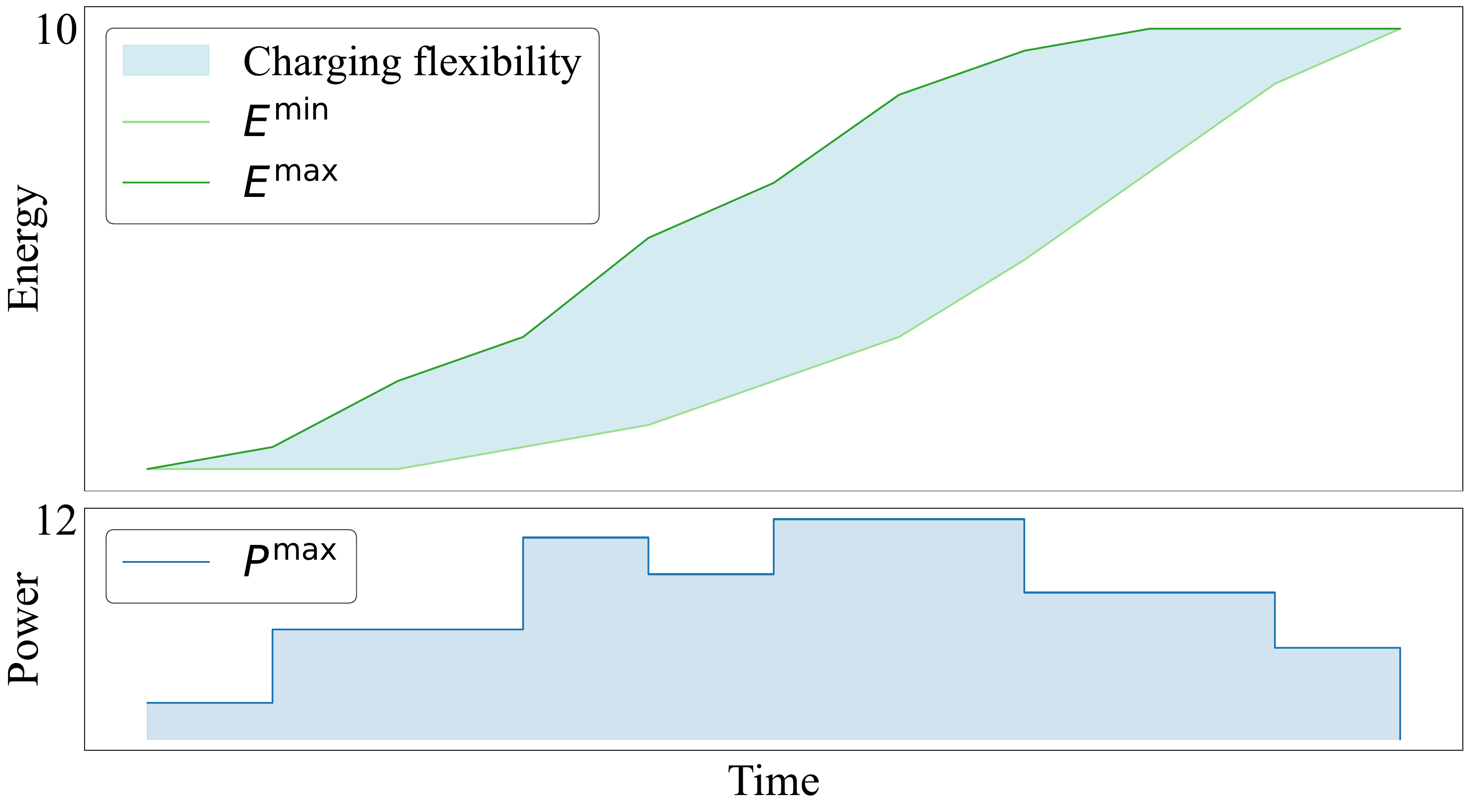}
  {\\ \footnotesize (b) Resulting virtual battery\par}

  \caption{Individual EV charging parameters aggregated into a virtual battery, adapted from~\cite{brinkel2023}.}
  \label{fig:virtualbattery}
\end{figure}

These envelopes can be recomputed from telemetry every 15~minutes based on arrivals, departures, and balance market activations. In practice, charging must stay within the power limit set by $P_t^{\max}$ and produce an energy trajectory that remains between $E_t^{\min}$ and $E_t^{\max}$. Although aggregation can, in theory, yield infeasible individual trajectories, quarter-hourly re-scheduling restores feasibility by adapting to the updated system state after each disaggregation. An example construction is shown in Fig.~\ref{fig:virtualbattery}.

\subsection{Risk-Aware Profit Maximisation}\label{sec:optModel}
We model the aggregator’s aim as choosing bids and an underlying charging plan that maximises revenue from mFRR activations while controlling downside risk from imbalance costs. Only market-side uncertainty is considered (system regulation state and the mFRR price), with EV flexibility treated as deterministic and hedged via the conservative policy in Section~\ref{sec:ev-uncertainty}. A market outcome scenario is denoted $\omega$.

\paragraph{Profit components}
Let the stochastic variable $\Pi$ denote total profit over the optimisation horizon. It combines
(i) mFRR activation revenues for {up} and {down} energy at the quarter-hour’s activation price,
(ii) the cost of energy used for charging sourced from the day-ahead market, and
(iii) imbalance settlement costs due to uninstructed deviations from the day-ahead commitment.

\paragraph{Day-ahead market bidding}
The position taken in the day-ahead market strongly affects mFRR bidding. 
One option is to operate under an \textit{independent} day-ahead commitment, where the day-ahead position is set from the day-ahead spot price and EV-demand forecasts, shifting charging to low-price hours. A more advanced approach is to \textit{co-optimise} day-ahead and mFRR participation, also including forecasts of mFRR scenarios as input to determine the day-ahead bid. Co-optimisation aligns the day-ahead position with ancillary service participation and increases expected profit. The value of this approach is assessed in this work.
\paragraph{Risk-aware objective}
The objective combines expected profit with downside risk through a parameter $\beta\in[0,1]$ and a confidence level $\alpha\in[0,1]$ (e.g., $0.95$):
\begin{equation*}
    \max \ (1-\beta)\,\mathbb{E}_\omega[\Pi]\;+\;\beta\,\mathrm{CVaR}_\alpha(\Pi).
\end{equation*}
Here, $\mathbb{E}_\omega[\Pi]$ is the expected profit across all scenarios $\omega$, and $\mathrm{CVaR}_\alpha(\Pi)$ is the average profit in the worst $(1-\alpha)$ fraction of scenarios (a tail average, not just a quantile). Setting $\beta=0$ gives a risk-neutral policy; larger $\beta$ places more weight on protecting the lower tail.

\paragraph{Fleet and market feasibility}
All decisions are constrained by the virtual-battery envelopes: the energy trajectory must remain between $E_t^{\min}$ and $E_t^{\max}$ and respect the instantaneous limit $P_t^{\max}$. Offered bid volumes cannot exceed available flexibility (including the safety buffer), activation in each scenario is given by cleared bids, and settlement follows the one-price imbalance rule. Integer-MW requirements for bids are enforced. Non-anticipativity constraints (same bid for same information set) ensure that decisions taken before uncertainty is revealed are identical across scenarios.

\subsection{Market Scenarios and Case Study Outline}\label{sec:scenCase}
\subsubsection{Scenario Generation}\label{sec:scengen}
The bidding framework optimises over a set $\Omega$ of market scenarios $\omega$. Since uncertainty in the mFRR market stems from both the system regulation state and the activation price, these components are modelled separately. 

For the regulation state, the system can be in \textit{up regulation}, \textit{down regulation}, or \textit{no regulation}. Transitions between states are represented by a non-time-homogeneous Markov chain, inspired by~\citep{olsson2008}. This allows the probability of moving to each state to depend on the current state and the time already spent in it. 

Conditional on the regulation state, an mFRR price is drawn: an up-regulation price is generated if and only if the system is up-regulating, and likewise for down regulation. Only the price \textit{premium} relative to the day-ahead market is explicitly modelled. These are defined as
\begin{equation*}
    \lambda_t^{\uparrow, \Delta} = \lambda_t^{\uparrow} - \lambda_t^\text{DA}, 
    \qquad
    \lambda_t^{\downarrow, \Delta} = \lambda_t^\text{DA} - \lambda_t^{\downarrow},   
\end{equation*}
where $\lambda_t^{\uparrow}$ and $\lambda_t^{\downarrow}$ denote mFRR~up and down prices, and $\lambda_t^{\text{DA}}$ the day-ahead price. Both premia are strictly positive by construction and are modelled using an autoregressive process with bootstrapped residuals and a mean-reversion term.

The time-series components are estimated on roughly five months of 15-minute data from price area SE3, starting on 19~March~2025 (when 15-minute pricing began). Since the initial weeks show unusually high price spikes, we trim the upper 1\% of the training data; the same trimming is applied when bootstrapping residuals for scenario generation.

\subsubsection{Case Study Setup}
In this paper, the optimiser is run once from a day-ahead vantage point to co-ordinate the day-ahead position and assess the expected value of mFRR participation. At day-ahead we have no certain information on which vehicles will be plugged in during delivery (vehicles connected now will not remain so the next day). Consequently, the fleet’s virtual-battery parameters are built as \textit{expected envelopes} informed by typical residential charging behaviour.

In lieu of operational history for this study, we approximate a representative fleet by sampling 1,000 long-duration residential sessions from the empirical distributions in Table~\ref{tab:EVparams}. Each sampled vehicle is described by $(T_v^\text{arr},\,T_v^\text{dep},\,E_v,\,P_v)$ with $P_v=11$~kW, and times rounded to 15~min. From this synthetic cohort we construct the virtual-battery envelopes $E_t^{\min}$, $E_t^{\max}$, and $P_t^{\max}$ over a full 24~h window from 13{:}00 to 13{:}00. To further hedge availability uncertainty, any bid must be backed by at least 10\% more instantaneous charging capacity (e.g., offering 1~MW requires more than 1.1~MW available).
\begin{table}[h]
    \centering
    \caption{Distribution of EV parameters used to generate synthetic fleet data, adopted from~\cite{huang2024} for residential, long-duration charging.}
    \label{tab:EVparams}  
    \begin{tabular}{l l l}
        \toprule
        \textbf{Parameter} & \textbf{Symbol} & \textbf{Distribution} \\
        \midrule
        Energy demand  & $E_v$          & $\operatorname{Lognormal}(2.6, 0.6^2)$ \\
        Arrival time   & $T_v^\text{arr}$ & $\operatorname{Normal}(17.1, 1.3^2)$ \\
        Departure time & $T_v^\text{dep}$ & $\operatorname{Normal}(8.9, 1.3^2)$ \\
        Charging power & $P_v$          & $11$~kW (fixed) \\
        \bottomrule
    \end{tabular}
\end{table}

To represent two stylised cases relevant for EV charging, we consider two observed SE3 day-ahead prices on:
\begin{enumerate}
    \item \textit{25--26 Mar 2025, 13{:}00--13{:}00} for a \textit{double-peak} profile with pronounced morning and evening peaks (approximately 137~EUR/MWh at 18{:}00-19{:}00 and 138~EUR/MWh at 07{:}00--08{:}00); and
    \item \textit{13--14 Jul 2025, 13{:}00--13{:}00} for a \textit{duck-curve} profile with a depressed midday price (minimum 0.15~EUR/MWh at 13{:}00-14{:}00) and higher evening prices.
\end{enumerate}

For each day-ahead spot price day, mFRR uncertainty (regulation state and activation premia) follows the framework in Section~\ref{sec:scengen}. The model returns a day-ahead charging baseline and indicative mFRR bids. Reported KPIs are expected profit and $\mathrm{CVaR}_{0.95}$; these should be interpreted as \textit{ex-ante} estimates under the scenario set. In practice, an aggregator would re-optimise every 15~minutes as telemetry arrives, rebuilding the virtual battery and potentially offering less conservative volumes. The implementation of this rolling set-up is outside the scope of this paper.

\section{Results and Analysis}\label{sec:results}
We evaluate the two representative day-ahead price profiles under both independent and co-optimised bidding strategies, using approximately 1,000 mFRR market scenarios. Table~\ref{tab:kpis} summarises financial metrics, a decomposition of the profit, and energy metrics across the different cases. The risk-aversion parameter is fixed at $\beta = 0.4$, based on a knee-of-the-curve calibration work.  

\begin{table}[h]
  \centering
  \caption{Key performance indicators by price profile and bidding mode.}
  \label{tab:kpis}
  \renewcommand{\arraystretch}{1.25}
  \resizebox{\columnwidth}{!}{%
    \begin{tabular}{@{} l *{4}{r} @{}} 
      \toprule
      \textbf{Metric} & \multicolumn{2}{c}{\textbf{Double-peak}} & \multicolumn{2}{c}{\textbf{Duck-curve}} \\
      \cmidrule(lr){2-3}\cmidrule(lr){4-5}
                      & Ind. & Co-opt. & Ind. & Co-opt. \\
      \midrule
      \textbf{Financial (EUR)} \\
      Total profit$^{*}$      & -111 & 135 & 356 & 673 \\
      $\mathrm{CVaR}_{0.95}^{*}$ & -646 & -425 & -133 & 271 \\
      \midrule
     \textbf{Profit decomposition (EUR)} \\
      Day-ahead contribution$^{\dagger}$           & -665 & -631 & -248 & -228 \\
      mFRR~up contribution$^{*}$       & 247 & 290 & 139& 186 \\
      mFRR~down contribution$^{*}$     & 77 & 157 & 427 & 622 \\
      Imbalance contribution$^{*}$     & 231 & 319 & 38 & 92 \\
      \midrule
      \textbf{Regulation / Imbalance (MWh)} \\
      Day-ahead volume$^{\dagger}$   & 14.2 & 10.2 & 14.2 & 8.0 \\
      mFRR~up$^{*}$     & 1.8 & 1.7 & 1.5 & 1.2 \\
      mFRR~down$^{*}$   & 8.8 & 12.2 & 8.5 & 12.7 \\
      Imbalance~up$^{*}$   & 7.0 & 7.0 & 7.1 & 5.9 \\
      Imbalance~down$^{*}$ & 0.1 & 0.4 & 0.1 & 0.5 \\
      \bottomrule
    \end{tabular}%
  }
  {\footnotesize $^{\dagger}$Deterministic, from day-ahead schedule\quad
   $^{*}$Expectation over scenario set}
\end{table}

\subsection{Charging Baseline, Offered Bids and Market Prices}
\begin{figure*}[t]
  \centering
  \begin{subfigure}[t]{0.485\textwidth}
    \centering
    \includegraphics[width=\textwidth]{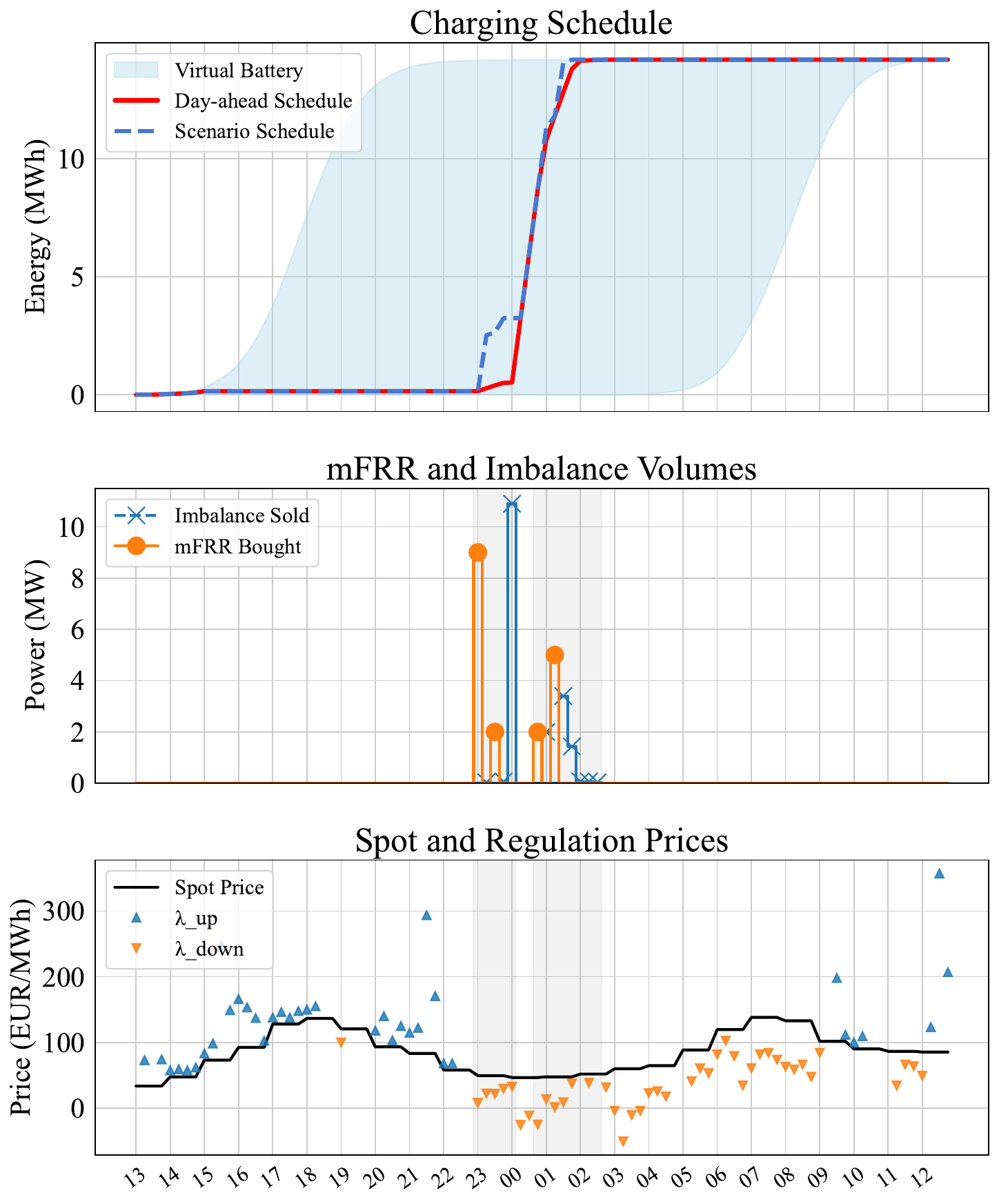}
    \subcaption{Independent case}
    \label{fig:doublepeak-ind}
  \end{subfigure}\hfill
  \begin{subfigure}[t]{0.485\textwidth}
    \centering
    \includegraphics[width=\textwidth]{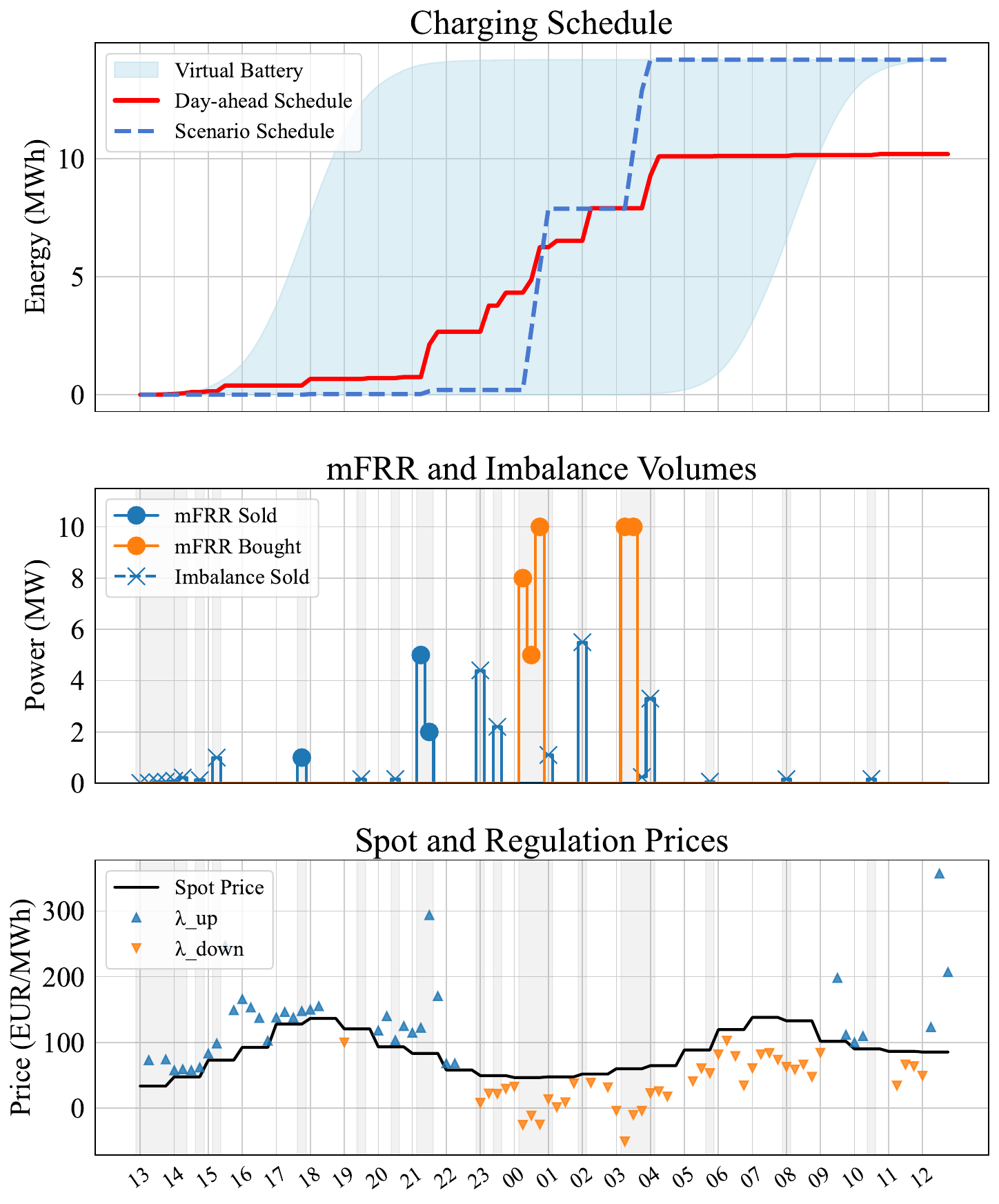}
    \subcaption{Co-optimised case}
    \label{fig:doublepeak-co}
  \end{subfigure}
  \caption{Double-peak day. Most-likely scenario shown in both panels for like-for-like comparison between (a) independent and (b) co-optimised bidding; panels plot charging vs.\ envelopes, mFRR/imbalance volumes, and prices.}
  \label{fig:bids-doublepeak}
\end{figure*}

\begin{figure*}[t]
  \centering
  \begin{subfigure}[t]{0.485\textwidth}
    \centering
    \includegraphics[width=\textwidth]{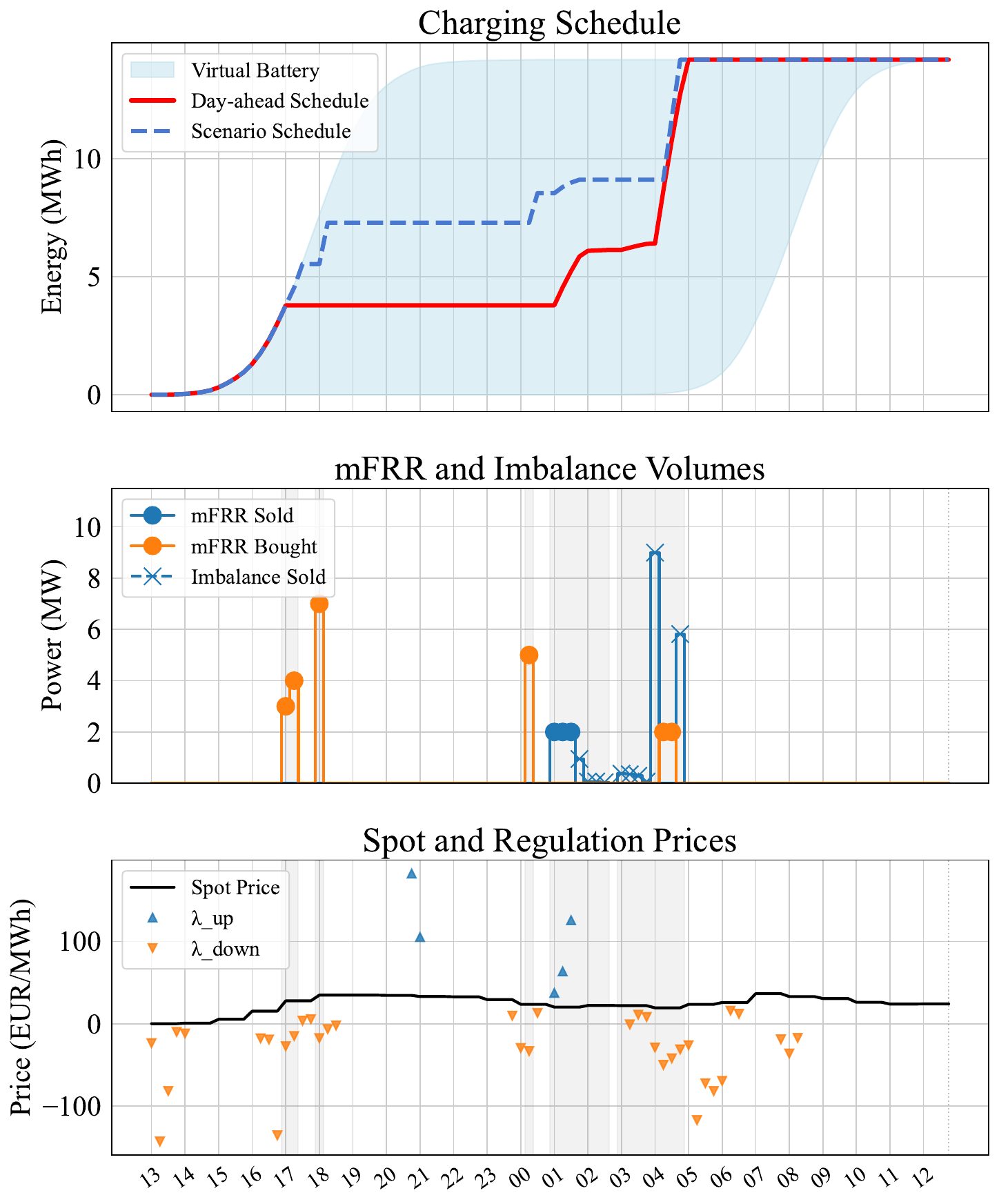}
    \subcaption{Independent case}
    \label{fig:duck-ind}
  \end{subfigure}\hfill
  \begin{subfigure}[t]{0.485\textwidth}
    \centering
    \includegraphics[width=\textwidth]{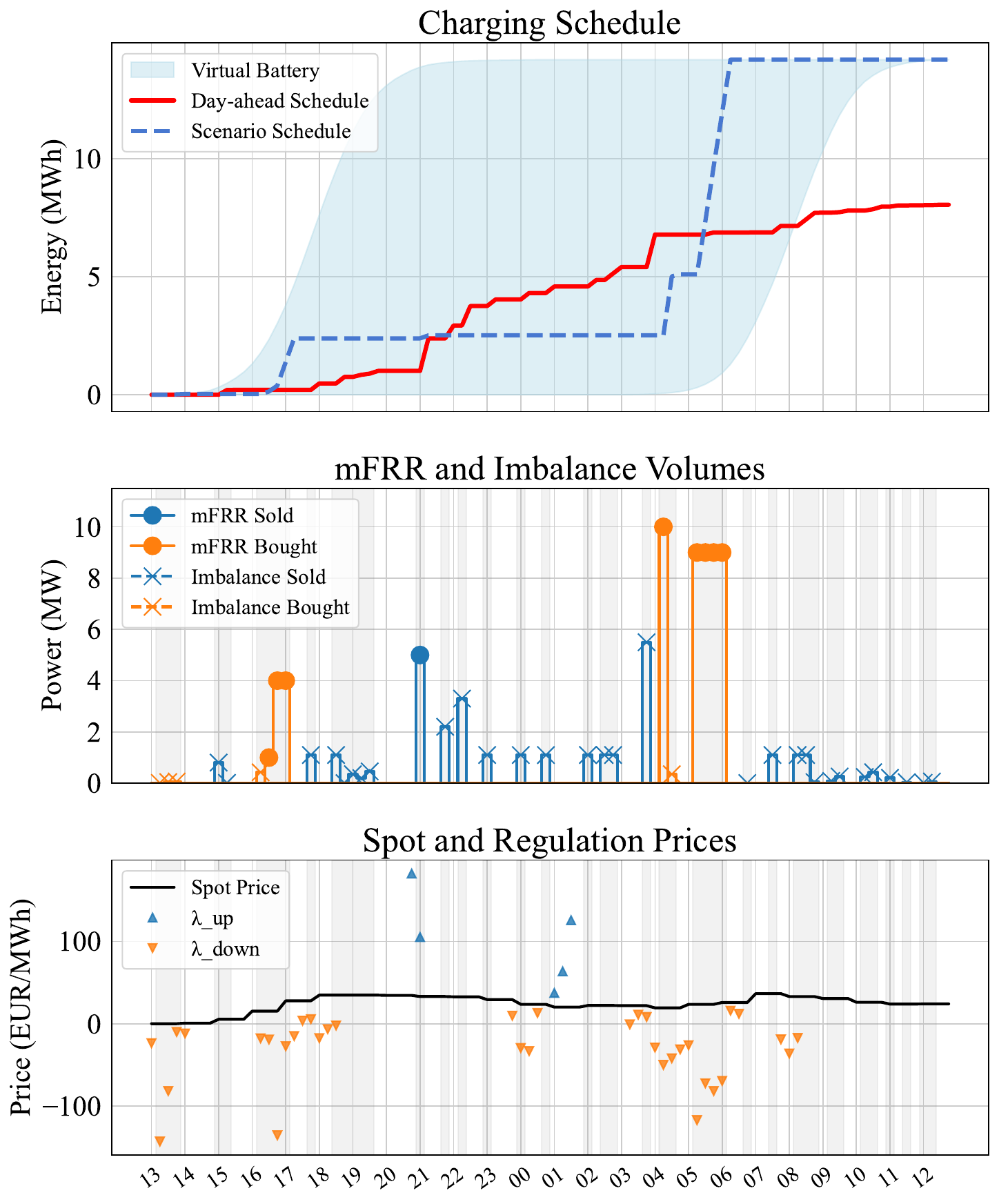}
    \subcaption{Co-optimised case}
    \label{fig:duck-co}
  \end{subfigure}
  \caption{Duck-curve day. Most-likely scenario shown in both panels for like-for-like comparison between (a) independent and (b) co-optimised bidding; panels plot charging vs.\ envelopes, mFRR/imbalance volumes, and prices.}
  \label{fig:bids-duck}
\end{figure*}

Fig.~\ref{fig:bids-doublepeak} presents the \textit{double-peak} day, followed by Fig.~\ref{fig:bids-duck} for the \textit{duck-curve} day. 
Each figure compares \textit{independent} and \textit{co-optimised} outcomes using the most likely scenario path, and the same path is replayed under both models for comparison. 
Each panel shows the charging trajectory against the virtual-battery envelopes, the quarter-hour activation and imbalance volumes, and the day-ahead and mFRR prices. 
These examples are intended to illustrate operational behaviour rather than to compare expected profits.

\subsection{Aggregate performance and profit decomposition}\label{sec:agg-perf}
Based on Table~\ref{tab:kpis}, two effects stand out. 
First, co-optimising the day-ahead commitment with mFRR participation improves both the expected outcome and the lower tail in both price profiles: for the double-peak day $\mathbb{E}[\Pi]$ moves from $-111$~EUR (independent) to $135$~EUR (co-optimised), and for the duck-curve day from $356$~EUR to $673$~EUR. The $\mathrm{CVaR}_{0.95}$ also improves (from $-646$ to $-425$~EUR in the double-peak case, and from $-133$ to $271$~EUR in the duck case).

Second, the co-optimised policy reduces reliance on the day-ahead schedule and shifts procurement towards mFRR~down. Rather than procuring the full 14.2~MWh required to charge the fleet in the day-ahead market, the co-optimised policy buys 10.2~MWh on the double-peak day and 8.0~MWh on the duck-curve day. Expected mFRR~down volume rises (from $8.8$ to $12.2$~MWh and from $8.5$ to $12.7$~MWh, respectively), while mFRR~up is slightly lower. In the duck-curve case, imbalance volume declines ($7.1$ to $5.9$~MWh), indicating less need to unwind the day-ahead position ex post.

The profit decomposition clarifies the mechanism. The largest uplift comes from mFRR~down revenue, which increases from $77$ to $157$~EUR on the double-peak day and from $427$ to $622$~EUR on the duck-curve day. This is enabled by a smaller day-ahead position, leaving headroom to meet charging needs via mFRR~down rather than pre-buying in the day-ahead market. mFRR~up rises more modestly (double-peak: $247$ to $290$~EUR; duck: $139$ to $186$~EUR).

The day-ahead contribution becomes less negative because fewer MWh are procured in the day-ahead market. Note, however, that the average day-ahead spot price paid per MWh is higher under co-optimisation, which is expected since the independent baseline picks hours to minimise day-ahead cost, while the co-optimised schedule sets the day-ahead position to support participation in mFRR. Consistent with this shift, the imbalance contribution also increases. This reflects higher \textit{price levels} when surplus day-ahead energy is unwound, not improved margins versus the original purchase. In other words, although imbalance volume may clear below the day-ahead price, they occur in higher-price hours than under the independent baseline, yielding larger gross imbalance revenue.

\section{Discussion and Outlook}\label{sec:discussion}
This section interprets the results in terms of bidding practice, provides a practical participation guide for aggregators, and outlines directions for future work. 

\subsection{Operational implications for bidding}\label{sec:bidding-implications}

\subsubsection{Roles of mFRR~up and mFRR~down}
Both products are remunerative in our experiments. On the double-peak day, mFRR~up edges out mFRR~down, while the reverse holds on the duck-curve day (Table~\ref{tab:kpis}). For EVs, however, mFRR~down is typically the better operational fit: it enables charging at very low net prices (often effectively being paid to charge), whereas mFRR~up defers energy intake and introduces a later cost of acquiring energy. 

In the Nordics, mFRR~up events tend to be rarer but carry higher premia, while mFRR~down events are more frequent with more modest premia; this frequency–premium trade-off makes mFRR~down the more reliable source of volume to plan against, with mFRR~up treated as opportunistic upside. This dynamic also concentrates downside risk on the energy make-up side: if insufficient energy is sourced via mFRR~down and the fleet must over-consume relative to its day-ahead plan, the deviation is settled at the quarter-hour's activation price. When that quarter-hour is in up regulation, high mFRR~up premia can make such unavoidable imbalances especially costly.

\subsubsection{How co-optimisation reshapes the day-ahead schedule}
The co-optimised model adapts the day-ahead position in two consistent ways.
(i) It buys less than the fleet’s full energy need day-ahead, relying on the high likelihood of sourcing part of the demand via mFRR~down. Because down events are more common than up events, a large day-ahead position would later force sell-offs at unfavourable prices; when co-optimising, the optimiser avoids purchasing that surplus in the first place.
(ii) It flattens the charging profile. Instead of a bang–bang plan with a few full-power hours, charging is spread across more hours. This increases the number of quarter-hours that satisfy the 1~MW minimum (with the 10\% availability buffer), improving eligibility for mFRR~up when it occurs. The flatter plan trades a few rare “jackpot” hours for many smaller, more frequent opportunities. With a less concentrated schedule, mFRR~up sales are less time-pressured and can be timed into better quarter-hours. Consequences include similar total day-ahead spot cost but less day-ahead energy (i.e., a slightly higher average day-ahead spot price per MWh), higher mFRR~up revenue, and lower aggregate volumes sold in mFRR~up and via imbalance, primarily because the day-ahead position is smaller. 

\subsection{Implementation Guide for EV Aggregator Participation in mFRR}
Long before operation, the aggregator needs a forecast of fleet flexibility to shape a day-ahead charging baseline that enables mFRR participation. A practical approach is to estimate a virtual battery, either by summing individual EV forecasts or by forecasting the aggregate directly, and co-optimise this baseline with day-ahead and mFRR price forecasts so headroom for down-regulation (and eligibility for up-regulation) is preserved. Because this baseline is not itself an mFRR \textit{energy activation} bid, the forecast should be representative rather than conservative. One exception is when also bidding in the mFRR \textit{capacity} market, where the forecast must be conservative to ensure deliverability if the bid clears.

During operation, real-time telemetry provides per-EV energy need, available charging power, and expected departure. With bids submitted 45 minutes before delivery, the aggregator must project the fleet state at delivery; a conservative policy counts only vehicles connected and not due to depart, while a less conservative policy forecasts arrivals and updates the virtual battery 45 minutes ahead, also accounting for incumbent bids that may clear before delivery. The optimiser should hedge for uncertainty in early disconnects or departures (e.g., require extra headroom to back offered volume).

\subsection{Future Research}
\subsubsection{From Day-Ahead to Rolling Optimisation}
The framework tested here is based on a once-per-day optimisation at the day-ahead stage, with profits evaluated in expectation. In practice, an aggregator would operate under a rolling horizon, updating the virtual battery and bids every 45~minutes as new telemetry and market information arrive. Moving to such a model predictive control set-up would test how forecast errors, arrivals, and early departures affect performance. It would also allow examination of how ex-ante profit expectations translate into realised outcomes under actual market results. Extending the framework in this direction is a natural next step.

\subsubsection{From V1G to V2G}
Bidirectional charging offers an attractive extension. By enabling discharge, the aggregator gains flexibility and can better manage exposure to high up-regulation prices. In the V1G setting, the main risk arises when extra charging creates imbalances in quarter-hours that unexpectedly clear with costly up-regulation. With V2G, this risk can be hedged by placing mFRR~up bids backed by potential discharge: if activated, the fleet avoids additional charging that quarter-hour and instead supplies energy at a premium. This turns an uncontrollable risk into an optional revenue stream, while still meeting energy requirements by charging later.

The virtual battery remains a suitable abstraction, but the optimisation model must be extended. Negative power should be allowed up to a discharging limit, and battery degradation represented, either through explicit costs or aggregate cycling constraints. Mapping individual degradation and switching limits into an aggregate model is not straightforward and requires further work. Future research should therefore focus on incorporating degradation and operational limits in a way that works at the aggregate level, while still reflecting physical constraints of individual vehicles.

\section{Conclusion}\label{sec:conclusion}
This paper examined the business case for EV fleet participation in the Nordic 15-minute mFRR market using a virtual-battery representation and a risk-aware optimisation of day-ahead and balancing decisions. We compared an \textit{independent} day-ahead baseline with a \textit{co-optimised} strategy across two representative price cases. Under the same scenario set and fixed risk attitude, co-optimisation improved both expected profit and the lower tail (CVaR).

The mechanism is straightforward. The co-optimised policy buys less energy day-ahead and relies more on mFRR~down, which is frequent and well suited to EV charging. It also flattens the charging plan, increasing eligibility for mFRR~up when it occurs. As a result, day-ahead volume falls below the fleet’s total energy need, mFRR~down revenues rise, and imbalance volume decreases relative to supplied balancing energy. Where imbalances remain, they are mainly a by-product of unwinding the day-ahead position.

The study is ex-ante and run once at day-ahead with conservative availability; it does not yet test rolling operation or bidirectional charging. Two immediate extensions follow: (i) a 45-minute rolling horizon implementation to assess realised performance under actual activations and arrivals, and (ii) a V2G extension allowing discharge-backed mFRR~up while accounting for aggregate degradation and switching limits. 

In summary, the results indicate that EV fleets can participate meaningfully in the Nordic 15-minute mFRR Energy Activation Market, and that aligning the day-ahead position with expected balancing opportunities is central to capturing value.

\section{References}  


\end{document}